\documentclass[conference]{IEEEtran}
\IEEEoverridecommandlockouts
\usepackage{cite}
\usepackage{amsmath,amssymb,amsfonts}
\usepackage{algorithmic}
\usepackage{graphicx}
\usepackage{textcomp}
\usepackage{xcolor}
\def\BibTeX{{\rm B\kern-.05em{\sc i\kern-.025em b}\kern-.08em
    T\kern-.1667em\lower.7ex\hbox{E}\kern-.125emX}}
\usepackage{hyperref}
\usepackage{multirow}
\usepackage{threeparttable}

\begin{document}

\title{\LARGE \textbf {TAXI: \underline{T}raveling Salesman Problem \underline{A}ccelerator with \underline{X}-bar-based \underline{I}sing Macros Powered by SOT-MRAMs and Hierarchical Clustering}
\thanks{This work is accepted by 62nd ACM/IEEE Design Automation Conference (DAC) 2025. 

This work was funded through the joint MOU between the Indiana Economic Development Corporation, Purdue University, imec with in-kind support from the Applied Research Institute. Yoo participated in this work as a postdoctoral researcher at imec \& Purdue University.}
}

%\and
%\and
%\and
%\and
%\and
 \author{
 Sangmin Yoo\textsuperscript{1,2,4},
 Amod Holla\textsuperscript{2},
 Sourav Sanyal\textsuperscript{2},
 Dong Eun Kim\textsuperscript{2},\\
 Francesca Iacopi\textsuperscript{1},
 Dwaipayan Biswas\textsuperscript{3},
 James Myers\textsuperscript{3},
 Kaushik Roy\textsuperscript{2}\\
 %\textit{imec USA}\textsuperscript{$\star$}, \textit{Purdue University}\textsuperscript{$\dagger$}, \textit{imec}\textsuperscript{$\ddagger$} \\
 \textsuperscript{1}\textit{imec USA} \& \textsuperscript{2}\textit{Purdue University}, West Lafayette, Indiana, USA, \textsuperscript{3}\textit{imec}, Leuven, Belgium,\\
 \textsuperscript{4}\textit{Oregon State University}, Corvallis, Oregon, USA
% \textsuperscript{$\dagger$}\textit{School of Electrical and Computer Engineering, Purdue University}, West Lafayette, Indiana, USA \\
% \textbf{Correspondence}:  Sangmin.Yoo@imec-int.com
}

\maketitle

\begin{abstract}
Ising solvers with hierarchical clustering have shown promise for large-scale Traveling Salesman Problems (TSPs), in terms of latency and energy. However, most of these methods still face unacceptable quality degradation as the problem size increases beyond a certain extent. Additionally, their hardware-agnostic adoptions limit their ability to fully exploit available hardware resources. 
In this work, we introduce TAXI -- an in-memory computing-based TSP accelerator with crossbar(Xbar)-based Ising macros. Each macro independently solves a TSP sub-problem, obtained by hierarchical clustering, without the need for any off-macro data movement, leading to massive parallelism. Within the macro, Spin-Orbit-Torque (SOT) devices serve as compact energy-efficient random number generators enabling rapid ``natural annealing". By leveraging hardware-algorithm co-design, TAXI offers improvements in solution quality, speed, and energy-efficiency on TSPs up to 85,900 cities (the largest TSPLIB instance). TAXI produces solutions that are only 22\% and 20\% longer than the Concorde solver's exact solution on 33,810 and 85,900 city TSPs, respectively. TAXI outperforms a current state-of-the-art clustering-based Ising solver, being 8$\times$ faster on average across 20 benchmark problems from TSPLib.
\end{abstract}


\section{Introduction}
Combinatorial optimization problems (CoPs), like the traveling salesman problem (TSP), are ubiquitous with applications in logistics, electronic design automation, genetics, and operations research \cite{CoPs,CoP-trend}. For TSPs, one of the most extensively studied CoPs, traditional methods~\cite{concorde} use heuristics like branch-and-bound and backtracking to find optimal solutions. However, such methods slow down exponentially as the problem size increases, due to exponential search space explosion. Additionally, traditional von Neumann architecture causes memory bottlenecks, rendering the TSP solvers inefficient for large instances due to repeated read-write operations during search-space exploration. This operating mismatch between the traditional TSP solver's execution pattern and von Neumann architecture's dataflow further slows down the optimization process, with the increase in problem size.

\begin{figure}[!t]
  \centering
  \includegraphics[width=\linewidth]{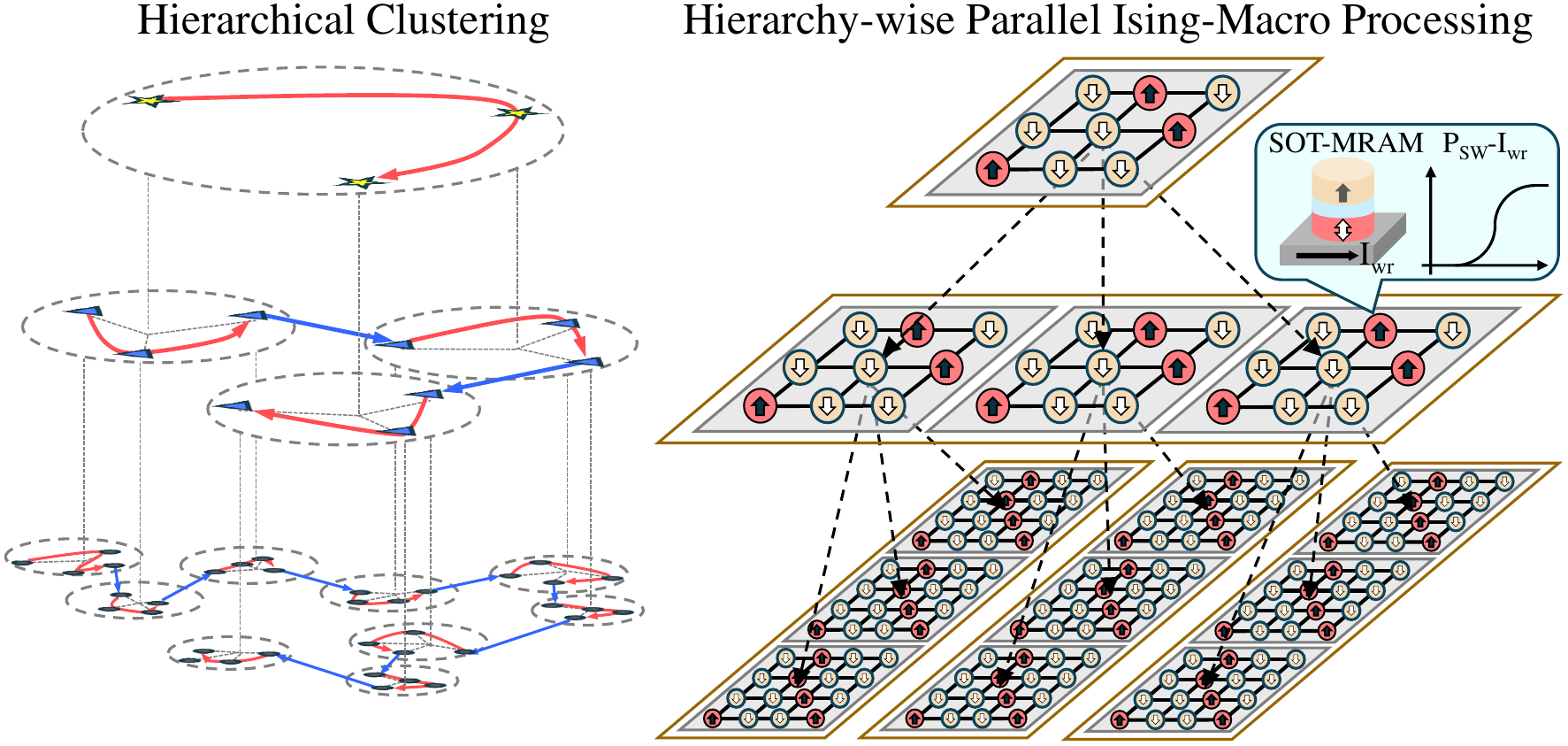}
  \vspace{-0.7cm}
  \caption{Hierarchical clustering and Ising macros for large-scale traveling salesman problems. The shortest route within each cluster (red in the left) is optimized by an Ising macro (each grey box in the right). The final solution is derived by merging inter-cluster (blue in the left) and intra-cluster routes.}
  \vspace{-0.6cm}
  \label{fig:Overview}
\end{figure}

Ising models, inspired by statistical physics, offer a promising alternative for solving small-scale TSPs by encoding combinatorial state-spaces onto 2D arrays of binary spin values, that naturally settle into minimal energy states. Small TSP instances can be efficiently mapped onto Ising crossbar (Xbar) arrays, potentially as hardware macros, enabling high-speed optimization. However, as the problem size grows, solution quality degrades due to the expanding search space and quadratic rise of spin interactions leading to a growing complexity of interconnects required to realize larger Ising Xbars. To address this issue, hierarchical clustering methods have been introduced~\cite{clustering-ising, neuro-ising, TSP-shimeng, TSP-shimeng2}, where large TSPs are decomposed into sub-problems that can be solved using smaller Ising models with acceptable quality. This work builds upon these clustering-based approaches, scaling further to tackle even larger problems with improved solution quality and further improvements in latency.

Typically, Ising solvers rely on stochastic processes to find near-optimal solutions. In conventional Complementary metal-oxide-semiconductor (CMOS)-based systems, random number generators (RNGs) are used to perform the stochastic switching of spins. However, these RNGs consume significant area and are too slow to meet the demands of rapid annealing, causing CMOS-based Ising solvers to be bulky with large area overheads and sluggish response-times~\cite{CMOS-RNG-DDlab,CMOS-RNG-JSSC12}. To overcome this, Spin-Orbit-Torque Magnetic Random Access Memory (SOT-MRAM)-based RNGs are considered in this work for Ising annealing, due to their energy-efficient and compact RNG implementation capability. In addition, in-memory computing (IMC) enables computations within memory arrays, reducing data movement and energy consumption. Xbar-based Ising primitives efficiently map TSP sub-problems onto hardware arrays \cite{TSP-shimeng,TSP-shimeng2}, making them suitable for realizing IMC macros. However, existing IMC-based Ising implementations still rely on off-macro memory accesses to store spin states.

To that effect, we propose TAXI -- a TSP accelerator that uses Xbar-based Ising primitives as IMC macros with minimal off-macro data movement. TAXI applies hierarchical clustering to decompose large TSPs into smaller sub-problems, which are then solved in parallel using dedicated Ising hardware units. This architecture ensures that each sub-problem is solved directly in memory.
This results in a hardware-algorithm co-design framework, where the algorithm (hierarchical clustering) and the architecture (spatially arranged Ising cores) are tightly integrated (See Fig \ref{fig:Overview}). The custom hardware is designed by considering the algorithm's hierarchical execution pattern, ensuring that the Ising computations align with the algorithm’s flow, enabling TAXI to maintain high solution quality as well as high speed even when problem size increases. Our contributions are as follows: \begin{itemize} 
\item A custom multiplication-and-accumulation (MAC)-based TSP solver using SOT-MRAMs and its stochastic switching to accelerate Ising annealing, overcoming the limitations of traditional CMOS-based RNGs (Section \ref{contri1}). 
\item A novel hierarchical clustering with fixed inter-cluster routes that reduces TSP scale, with independently invoked Ising sub-solvers to enhance parallelization and sub-solution quality, enabling near-optimal solutions for large-scale TSPs (Section \ref{contri2}). 
\item A crossbar (Xbar)-based parallel architecture that supports in-macro Ising computing through hardware-algorithm co-design, mapping sub-problems onto Xbar arrays, and aligning the custom hardware with the algorithm to improve scalability (in terms of latency and energy) for large-scale TSPs (Section \ref{contri3}).
\end{itemize} 
Our experimental results (Section \ref{sec:results}) demonstrate that TAXI is 8$\times$ faster on average across 20 benchmark problems (with up to 85,900 cities) from TSPLib \cite{tsplib} compared to a state-of-the-art hierarchical Ising solver~\cite{neuro-ising}. TAXI generates 3\% shorter traveling route on 33,810 cities over method \cite{TSP-shimeng2} and 31\% shorter route on 85,900 cities over method \cite{neuro-ising} -- two recent clustering-based Ising works which demonstrate the best solutions on the respective problems, so far.

\section{BACKGROUND AND RELATED WORKS}

\subsection{\textbf{Ising Solver}}\label{sec:Background-Ising-Solver}

In Ising model, Hamiltonian (energy) of the total system ($H_{total}$) and each spin (element) ($H_{i}$) are defined as below:
\begin{equation}
\label{equ:hamiltonian_total}
H_{total}=-\sum_{i}\sum_{j}J_{ij}\sigma_{i}\sigma_{j}-\sum_{i}h_{i}\sigma_{i} 
\end{equation}
\begin{equation}
\label{equ:hamiltonian_i}
H_{i}=\sum_{j}J_{ij}\sigma_{j}+h_{i} 
\end{equation}
where $J_{ij}$ represents interaction between spins, $\sigma_{i}$ and $\sigma_{j}$, while $h_{i}$ denotes external field applied to spins. Equation~\ref{equ:hamiltonian_total} can be reformulated by equation~\ref{equ:hamiltonian_i} as below:
\begin{equation}
\label{equ:hamiltonian_total_i}
H_{total}=-\sum_{i}\left(\sum_{j}J_{ij}\sigma_{j}+h_{i}\right)\sigma_{i}
\end{equation}
According to the rule that spin state, +1 or -1, follows the polarity of $H_{i}$, greater than 0 or not, respectively, $\Delta H_{total}$ is \textit{always negative}. It means every spin update minimizes $H_{total}$, approaching an optimal solution in an unsupervised manner.

Inspired by the nature, Ising solvers have been proposed to naturally find the optimal solution to problems that can be properly encoded in Ising format. For Ising acceleration, MAC operation based on a memory Xbar has been deployed~\cite{Cai2020,TSP-shimeng2} for $\sum_{j}J_{ij}\sigma_{j}$ in equation~\ref{equ:hamiltonian_i}; Spin states ($\sigma_{j}$) are applied to rows and multiplied with $J_{ij}$. Then, $H_{i}$ is derived.

\begin{figure}[!h]
  \vspace{-0.3cm}
  \centering
  \includegraphics[width=0.7\linewidth]{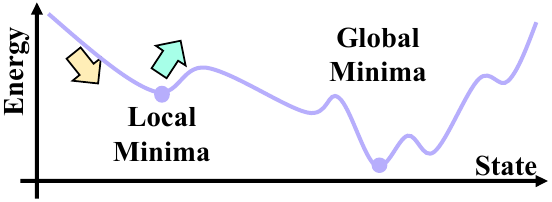}
  \vspace{-0.4cm}
  \caption{Non-monotonic energy search space. Energy minimization (yellow) and Stochastic update (green) jointly find the global minima by enabling descending the energy landscape and escaping from local minimas, respectively.}
  \vspace{-0.3cm}
  \label{fig:E-Space}
\end{figure}

However, the always-descending energy update is insufficient to get the optimal solution when problems are complicated. The complication makes the energy search space of Ising problem non-monotonic as depicted in (Fig~\ref{fig:E-Space}), causing the Ising solver to get stuck in local minimas. To climb up the energy hill and reach global minima which is optimal solution, Ising solver needs the power to violate the always-descending nature. For the violation, stochastic switching of spin disregarding equation~\ref{equ:hamiltonian_total_i} is necessary.

\subsection{\textbf{Source of Stochasticity}}
CMOS-based random number generator (RNG)~\cite{SRAM-pseudo,pingpong-ising,TSP-shimeng2} and emerging electronic devices, such as resistive random access memory (RRAM)~\cite{Cai2020}, silicon-oxide-nitride-silicon (SONS) FinFET~\cite{TSP-shimeng}, and spin-transfer torque MRAM~\cite{TSP-STT-Roy, Fukushima_2014, MTJ_RNG, MTJ_RNG2, pbits} have been deployed as the source of stochasticity.\par

RRAM and SONS FinFET offer stochasticity through their inherent thermal and temporal noise when used for a Xbar array for MAC operation~\cite{Cai2020,TSP-shimeng}. Despite their better chip density, %Even though such devices offer better chip density, 
the intrinsic noise becomes uncontrollable when the Xbar grows~\cite{Cai2020}. Magnetic tunnel junction (MTJ)-based RNGs based on low barrier nanomagnets and telegraphic switching \cite{Fukushima_2014, MTJ_RNG, MTJ_RNG2, pbits} and CMOS-based RNGs~\cite{CMOS-RNG-JSSC12, CMOS-RNG-DDlab} can be placed in the peripheral circuitry, which are not scaled up by Xbar size. However, current MTJ-based implementations require very low barrier nanomagnets ($\Delta \sim0kT$) for truly high-speed RNG ($>$1Gb/s) and high-frequency sense circuitry to detect rapid switching of the device, while CMOS-based RNGs are slower ($<$2400Mb/s)~\cite{CMOS-RNG-JSSC12} and take much more area ($>375\mu m^{2}$)~\cite{CMOS-RNG-DDlab}.

\subsection{\textbf{Ising Solvers for Large-scale TSP}}

Large-scale TSPs are hard for Ising solvers due to the quadratically growing number of connections as TSP size increases. Hierarchical Vertex Clustering (HVC)~\cite{clustering-ising} proposed hierarchical clustering to reduce the number of connections. Despite the reduction, it still sparsely maps all the clustered problems to a single matrix, which leaves the connection problem unsolved when the problem size is beyond a certain level. Neuro-Ising~\cite{neuro-ising} proposed a hybrid clustering-and-graph-neural network manner but shows performance degradation as the problem size increases.
In-memory annealer (IMA)~\cite{TSP-shimeng} and its follow-up, digital compute-in-memory annealer (CIMA) \cite{TSP-shimeng2}, based on the hierarchical clustering achieves acceleration by simultaneously optimizing non-adjacent clusters. However, they store the spin states outside macros, causing slow-down.

\section{SOT-MRAM-BASED TSP SOLVER}
\label{contri1}
\subsection{\textbf{MAC-based Energy Minimization for TSP}} \label{sec:Method-MAC-based-Hmin}

TAXI leverages the Xbar-based MAC operation to minimize $H_{total}$. To achieve the objective of TSP, finding the shortest route to visit all cities only once, information on the distance between cities and the visiting order initialized by input order are used. The distance is reformulated as follows:
\begin{equation}
\label{equ:distance-mapping}
W_{D}(A,B)=\frac{D_{min}}{D_{A-B}}\times B_{Precision}
\end{equation}
where $W_{D}(A,B)$ denotes a conductance value to map on a cross-point of Xbar, while $D_{A-B}$ and $D_{min}$ represent the distance between cities $A$ and $B$ and the minimum distance among all paths, respectively, and $B_{Precision}$ is the bit-precision that can be offered by the hardware.

\begin{figure}[!h]
  \vspace{-0.3cm}
  \centering
  \includegraphics[width=0.9\linewidth]{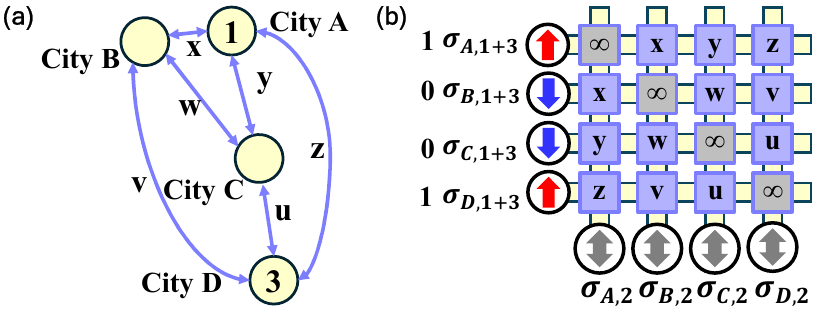}
  \vspace{-0.3cm}
  \caption{(a) TSP with 4 cities. $u-x$ represent distances and numbers in a circle denote visiting orders. (b) Distance matrix mapped on a crossbar. Distances reformulated to $W_{D}$ are programmed to crosspoints in resistance.}
  \label{fig:mapping}
  \vspace{-0.2cm}
\end{figure}

\begin{figure*}[t]
  \centering
  \includegraphics[width=0.85\linewidth]{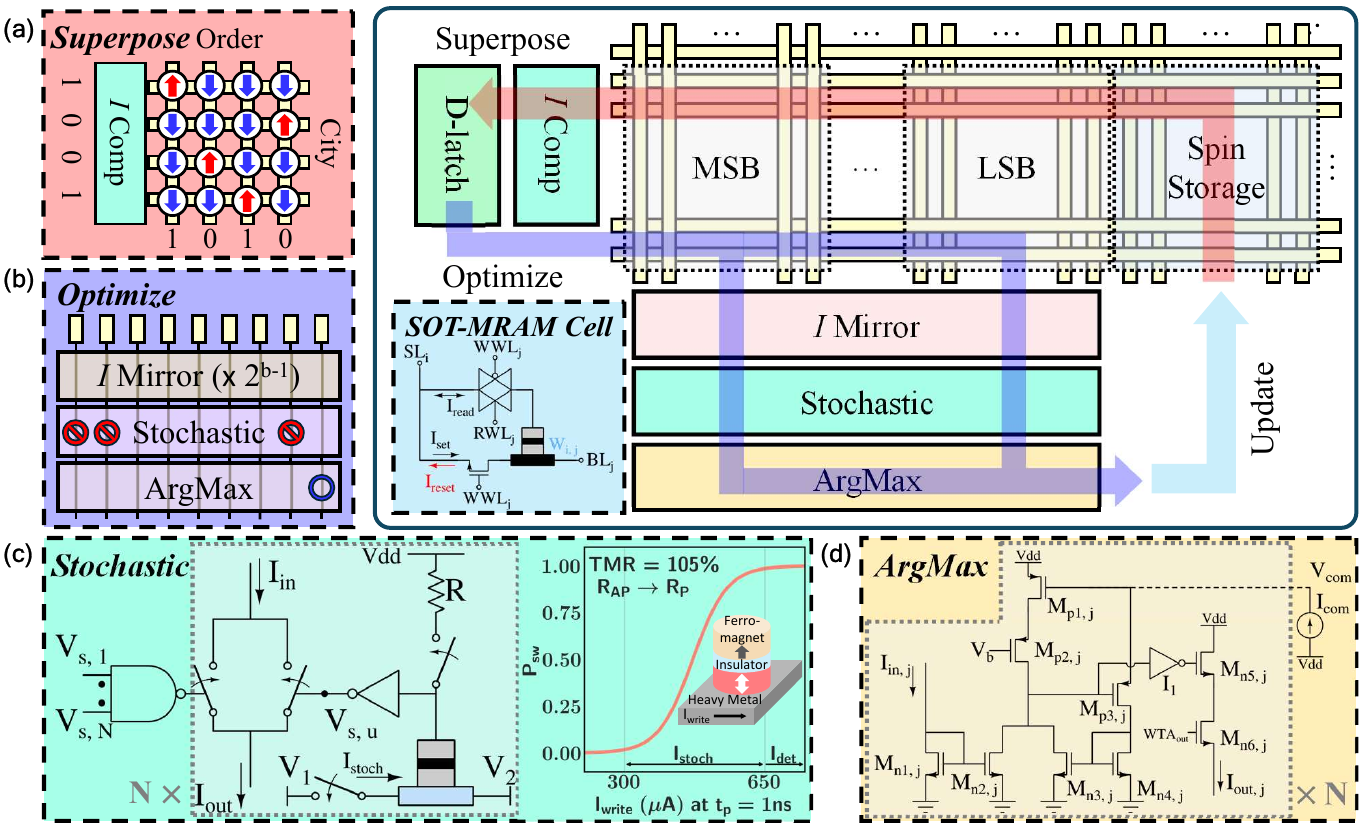}
  \vspace{-0.3cm}
  \caption{Floor plan of an Ising macro. Arrows represent data flow across the macro. Insets (a-d) illustrate components in the macro. (a) Superposition of vectors to optimization. Red and blue arrows represent up- and down-spin stored in spin storage, respectively. (b) Optimization by crossbar array as the distance matrix and following peripheral circuits. (c) SOT-MRAMs in the stochastic circuit pass or prevent current from the crossbar following its stochastic switching. Inset displays the probability of switching $R_{AP}$ to $R_{P}$ of the chosen SOT device~\cite{IMEC-SOT-MRAM2022}. $I_{\text{stoch}}$ and $I_{\text{det}}$ denote the current ranges for stochastic and deterministic operations, respectively. (d) ArgMax circuit picks a city to visit by choosing the largest current.}
  \label{fig:Ising-Macro}
  \vspace{-0.4cm}
\end{figure*}

Fig~\ref{fig:mapping} displays how the $W_{D}$ is mapped on the Xbar and used to choose a city to visit at order $\textit{i}$ which results in the shortest route. For the optimization, cities visited at the previous ($\textit{i}$-1) and the next ($\textit{i}$+1) orders are considered because the total traveling distance is decided by both. The visiting information ($\sigma_{A,i}$) to a specific city ($\textit{A}$) at a specific order ($\textit{i}$) is represented by a binary spin, 1 or 0, following quadratic unconstrained binary optimization (QUBO) and Ising equivalence~\cite{QUBO-Ising-Equivalence}. $\sigma_{A,1}$ and $\sigma_{D,3}$ are 1 because they are visited at order 1 and 3, respectively. With the fact that the distance information is static, we superpose the vector representing visiting information of order $\textit{i}$-1 ($\sigma_{A-D,1}$) and $\textit{i}$+1 ($\sigma_{A-D,3}$), as shown in Fig~\ref{fig:mapping}b. It reduces the crossbar array sizes, leading to better area-efficiency and less involvement of non-ideality.

In this setup, the current flowing through each column ($\sigma_{A-D,2}$) following Ohm's law and Kirchhoff's current law represents relative distance when the city is chosen for the order. It can be expressed as:
\begin{equation}
\label{equ:hamiltonian_tsp}
D_{x,i}=\sum_{k\neq x}W_{D}(x,k)(\sigma_{k,i-1}+\sigma_{k,i+1}) 
\end{equation}
According to equations~\ref{equ:distance-mapping} and~\ref{equ:hamiltonian_tsp}, the current flowing through the 3rd column becomes the largest due to shorter distance (lower resistance), making $\sigma_{C,2}$ selected as up-spin.

\subsection{\textbf{Stochasticity-Involved Decision Making} }\label{sec:SID}

In TAXI, stochasticity is involved in the choice of the largest current via a stochastically produced binary vector as big as the problem size. Only cities corresponding to 1s in the vector are considered for the choice. It offers the power to violate the always-descending nature, mentioned in Section~\ref{sec:Background-Ising-Solver}. The visiting order is updated by the newly chosen city.

\subsection{\textbf{Ising Macro as a TSP Solver}}

A Xbar array with customized peripheral circuitry is leveraged as an Ising macro to solve a TSP in the manner of the aforementioned algorithms. Fig~\ref{fig:Ising-Macro} shows the system-level layout of the Xbar-based Ising macro along with the associated peripherals inset in Fig~\ref{fig:Ising-Macro}c,d. The unit cell of the Xbar array is a 3T-1M SOT-MRAM unit, with a transmission gate to control the read current, and 1 transistor to control the write current. The transmission gate was chosen owing to the need for accurate bi-directional reads.

The Xbar is divided into $B+1$ partitions, where $B$ is the chosen bit-precision for the $W_{D}$. The first $B$ partitions contain the $W_{D}$, while the last partition stores the spin information ($\sigma_{k,i}$) as the solution of the given TSP. To minimize effect from non-ideality, such as wire resistance, higher significant bit is stored closer to the left end. The rows and columns of the spin storage (SS) represent cities and visiting orders. If a specific city (A) is chosen at a specific order (1), a SOT-MRAM that is located in the first row and the first column in the SS is programmed in the low resistance state (LRS), while the others on the same column are in the high resistance state (HRS). This implies that the size of the Xbar required for a TSP problem size of $N$ is $N\times N\cdot (B+1)$.

\subsubsection{\textbf{Superpose}} \label{sec:superpose}

The superposition of vectors representing visiting information, discussed in Section~\ref{sec:Method-MAC-based-Hmin} and Fig~\ref{fig:mapping}, is executed by MAC operation using the SS, as illustrated in Fig~\ref{fig:Ising-Macro}a. In the case of Fig~\ref{fig:mapping}, the first and the third columns are activated to recall visiting information at order 1 and 3. Then, the Xbar natively derives the superposed vector in current at the end of the rows, while a following current comparator\cite{current-comparator} translates it in binary and a D-latch stores the binary vector. This flow is depicted by the red arrow in Fig~\ref{fig:Ising-Macro}.

\subsubsection{\textbf{Calculate Distance}}
The stored vector in the D-Latch is fed back to the rows, generating currents by executing equation~\ref{equ:hamiltonian_tsp}, based on the $W_{D}$ stored in the first $B$ partitions of the Xbar. Only columns in the first $B$ partitions are activated, with columns in the SS deactivated. The currents from the Xbar flow into current mirrors, scaled by 2$^{b-1}$ depending on the bit-significance ($b$) of the partition, as shown in Fig.~\ref{fig:Ising-Macro}b.

\subsubsection{\textbf{Stochastic Binary Vector} }\label{sec:stochastic}

The stochastic binary vector mentioned in Section~\ref{sec:SID} is generated by SOT-MRAM's stochastic switching-based circuit consisting of $N$ identical units shown in Fig~\ref{fig:Ising-Macro}c.
SOT-MRAM has two ferromagnetic layers (one with a fixed magnetic moment direction, while another is free to rotate) separated by a thin insulator. Its resistance is determined by how many electrons can tunnel through this MTJ, which depends on whether the magnetic moment of the two ferromagnetic layers is aligned in parallel ($R_{P}$) or anti-parallel ($R_{AP}$). This alignment can change due to the Spin Hall effect induced by current flowing through a heavy metal layer under the ferromagnet~\cite{SOT_ref1}. The stochastic switching that TAXI leverages is based on this dynamic. Depending on the amount of current, the probability of switching ($P_{sw}$) between $R_{P}$ and $R_{AP}$ can be controlled, as shown in Fig~\ref{fig:Ising-Macro}c~\cite{IMEC-SOT-MRAM2022}. By controlling the current that derives a specific $P_{sw}$ (300$\mu$A --- 650$\mu$A), TAXI can control the expected number of 1s in the stochastic binary vector. SOT-MRAMs in the Xbar are operated only in the deterministic regime ($>$650$\mu$A).

The stochastic circuit (Fig.~\ref{fig:Ising-Macro}c) controls the flow of the preceding current mirror outputs into the next stage upon the stochastic switching. For a given unit, if the SOT device switches, the voltage across the divider circuit becomes lower than units whose SOT device has not switched. The output of the inverter is therefore high, resulting in the unit allowing the input current to pass through. The circuit allows the current only through the columns ($\mathcal{S}$) where the SOT device has stochastically switched for a given iteration to pass through. In case $\mathcal{S} = \varnothing$, currents through all columns are allowed to pass because the NAND gate output is high for every unit.

\subsubsection{\textbf{Choosing the Largest Current}}

The largest current is selected by the ArgMax circuit shown in Fig~\ref{fig:Ising-Macro}d. It processes a vector of input currents and outputs a vector where only the index `$i$' corresponding to the largest input current remains nonzero. The winner current is set to the minimum required to switch a SOT-MRAM device in the deterministic regime. The ArgMax circuit is based on Lazzaro's winner-take-all circuit~\cite{NIPS1988_a8f15eda}, enhanced with a cascaded transistor ($M_{p2, j}$) in the input branch to increase output resistance~\cite{cascode_wta}, and a current mirror ($M_{n2, j}$, $M_{n3, j}$) to boost feedback, significantly improving resolution and speed~\cite{wta_current_mirror}.

\subsubsection{\textbf{Update Spin Storage}} \label{sec:update-SS}

The one-hot encoded current vector derived by the ArgMax circuit is directly used to update the spin storage, as illustrated by the cyan arrow in Fig~\ref{fig:Ising-Macro}. Before the update operation, all the devices on the column representing the order which is being optimized are reset to HRS which represents 0. Then, the output current vector of the ArgMax circuit is applied to write the column.

\begin{figure*}[!t]
  \centering
  \includegraphics[width=\linewidth]{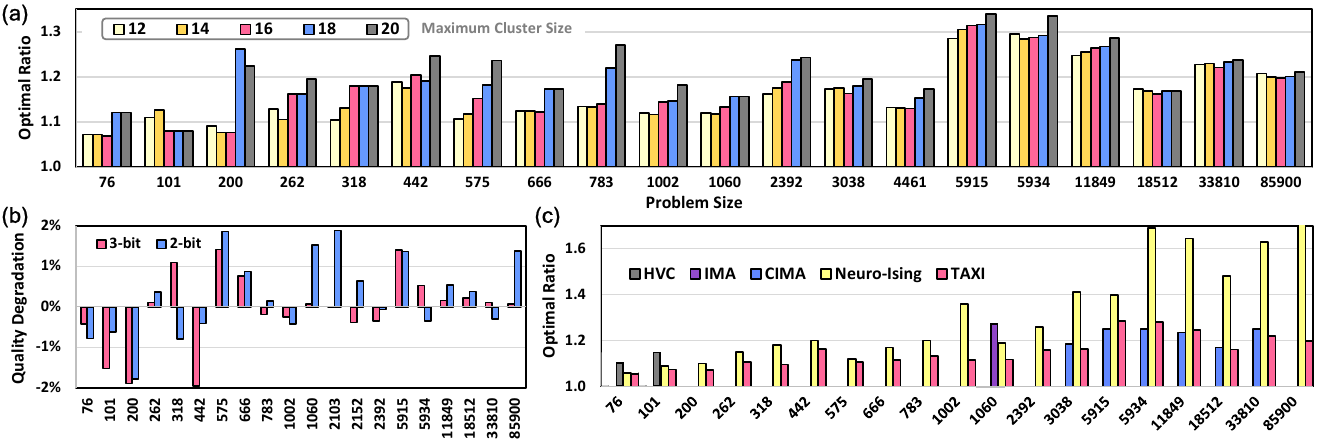}
  \vspace{-0.8cm}
  \caption{(a) Optimal ratio depending on the maximum cluster size in 4-bit precision. (b) Solution quality degradation when the bit precision changes from 4-bit to lower-bit options. The positive number represents degradation. (c) Comparison of solution optimality. Data of TAXI with 4-bit precision and 12 cluster size are presented. Data of other Ising solvers are adapted from~\cite{neuro-ising,TSP-shimeng,TSP-shimeng2}.}
  \label{fig:Results}
  \vspace{-0.3cm}
\end{figure*}

\subsubsection{\textbf{Annealing}}

The processes described in sections~\ref{sec:superpose} -~\ref{sec:update-SS} are repeated for the pre-designated number of iterations from the first to the last order by activating different columns in the SS depending on the visiting order to optimize. After the whole iteration, the spin information stored in the SS is retrieved as the final solution of the given TSP. To optimize the solution, stochasticity should decrease during iterations, allowing the Ising system to remain at the global minimum as the solution improves. Otherwise, the system may escape the global minimum and end up with a worse solution. The solution is optimized the best when the stochasticity decreases slowly. Optimization is however faster in the early stages making the non-linear decrease of stochasticity appealing.

For the non-linearity, TAXI utilizes the native sigmoidal switching probability vs $I_{write}$ characteristic, as shown in Fig~\ref{fig:Ising-Macro}c. The non-linearity can enable a relatively rapid decrease in stochasticity at the earlier stage, leading to shorter overall latency, while keeping the advantage of slow decrease at the later stage. To leverage the native sigmoidal decrease, the $I_{write}$ is initialized to 420$\mu$A (corresponding switching probability, 20\%) and linearly decreased by 50nA after every iteration. Once the $I_{write}$ becomes 353$\mu$A (corresponding probability, 1\%), the Ising solver stops, then, the solution stored in the SS is taken as the final output of the given TSP.
% \vspace{-1.5mm}

\section{HIERARCHICAL CLUSTERING}
\label{contri2}

\subsubsection{\textbf{Locally Solve Global Problems}}
We adopt a hierarchical structure similar to HVC~\cite{clustering-ising} for clustering, executed in a bottom-up fashion. After clustering all cities at level $i$, the cluster centroids form level $i+1$. This process repeats until the number of centroids at a level is less than the maximum TSP size confidently solvable by an Ising macro. The number of clusters at each level is determined by the maximum size.

\subsubsection{\textbf{Parallel Processing}}

After the hierarchical structure is defined, the TSP is solved in a top-down manner. Once the topmost TSP (level $L$) is solved, the solution, which is the visiting order of clusters at level $L-1$, is determined. Unlike HVC, which sparsely maps all clusters to a single Xbar array to optimize routes both within and between clusters, TAXI fixes the first and last cities of sub-problems first to minimize the inter-cluster route. This fixation ensures that sub-problem solutions do not compromise the inter-cluster distance. The first and last cities of each cluster are determined by identifying the closest city pairs between neighboring clusters, enabling parallel solving of as many clusters in a level as there are Ising macros in a chip. The entire process concludes when all sub-problems in the lowest level are solved.

\subsubsection{\textbf{Agglomerative Clustering}}

TAXI leverages Agglomerative clustering to achieve the hierarchical clustering for its robustness to outliers, instead of K-means clusters adopted by others~\cite{clustering-ising,TSP-shimeng, TSP-shimeng2, neuro-ising}. It is a type of hierarchical clustering that merges clusters based on an inter-cluster similarity metric until the desired number of clusters is obtained~\cite{Agglomerative-clustering}. With the Ward linkage, the squared distance or variance within a cluster is minimized~\cite{ward_linkage}. Even though K-means clustering also aims to minimize the intra-cluster variance, they usually lead to nearly spherical or regular clusters~\cite{JAIN2010651}, while agglomerative clustering can produce compact irregular clusters.

\section{X-BAR-BASED PARALLEL ARCHITECTURE}
\label{contri3}
Hierarchical clustered Ising macros are mapped to the Xbar architecture to enable in-macro Ising computing through hardware-algorithm co-design, mapping sub-problems onto Xbar arrays, and aligning the custom hardware with the algorithm to improve scalability for large-scale TSPs. 
We instrumented the PUMA architecture~\cite{PUMA} to perform the mapping and evaluated the latency and energy consumption from data movement between off-chip memory and the proposed Ising macros executing parallel Ising computations. PUMA is a cycle-accurate simulator based on an IMC spatial architecture, structured with a hierarchy of chip, tile, core, and MVMUs. The simulator’s compiler generates instructions using PUMA ISA, and the simulator executes them to assess latency and energy. We replaced PUMA's ReRAM-based MVMUs with our Ising macro and scaled all elements from PUMA's 32nm to a 65nm node. See Section \ref{sec:architecture-results} for more details.

\section{EVALUATION}
\label{sec:results}

\subsection{\textbf{Results on Multiple TSPs}} \label{sec:algorithmic_sim}

The performance of TAXI with varied maximum cluster sizes is simulated in C++, following the Ising layer presented in~\cite{neuro-ising}. For realistic simulation, ON/OFF resistance of SOT-MRAM and transistors and wire resistance are considered~\cite{IMEC-SOT-MRAM2022}. Fig~\ref{fig:Results}a displays the optimal ratio which is derived by dividing a solution of Ising solver with exact solution~\cite{concorde-benchmark}.
A smaller cluster (sub-problem) size leads to a better solution quality in most cases as Ising solvers usually show better results on smaller problems.
With the problem size of 12, we tested different bit precisions for the $W_{D}$.

Fig~\ref{fig:Results}b presents solution quality degradation when the bit precision changes from 4-bit to 3- or 2-bit. It shows that the performance is maintained in the 2\% range. We attribute this fluctuation to variation of both $W_{D}$ precision and non-ideality arising from Xbar size change. From the energy-efficiency perspective, the lower bit-precision is beneficial. Depending on the priority between the power budget and the solution quality, TAXI can be reconfigured. The comparison will be discussed in the following section~\ref{sec:architecture-results}.
TAXI with a clustering size of 12 and 4-bit precision is benchmarked with existing Ising solvers for TSPs in Fig~\ref{fig:Results}c. TAXI outperforms the others~\cite{neuro-ising,TSP-shimeng,TSP-shimeng2} in most cases including three of the largest TSPs, 18512, 33810, and 85900.

\subsection{\textbf{Circuit Simulation}}
An entire Ising macro shown in Fig~\ref{fig:Ising-Macro} is simulated using the \texttt{Cadence}\textsuperscript{\textregistered}  \texttt{Spectre}\textsuperscript{\textregistered} simulator in the TSMC 65nm technology node. Verilog-A behavioral modeling was performed to fit the characteristics of the considered SOT device in section \ref{sec:stochastic}. We choose a TSP problem size of 12. We run the simulation for 1 complete iteration including the superposition, optimization, and spin-storage updates. The pre-layout circuit simulation results are presented in Table~\ref{tab:circuit}. The higher-bit precision causes a larger area and higher energy consumption.

\begin{figure}[!t]
  \centering
  \includegraphics[width=\linewidth]{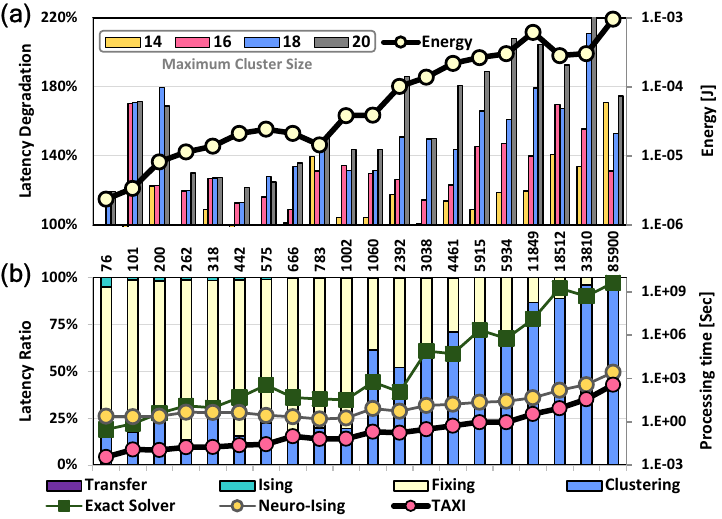}
  \vspace{-0.6cm}
  \caption{(a) Latency (bar) and energy (line) arising from Ising macros and data transfer within the architecture depending on the maximum cluster size in 4-bit precision. Energy based on a problem size of 12 and 2-bit precision is representatively displayed. (b) Total Latency including clustering, fixing, Ising processing, and data transfer illustrated in lines (right y-axis). Bars represent the ratio of each component contributing to the total latency (left y-axis).}
  \label{fig:puma_result}
  \vspace{-0.3cm}
\end{figure}

\renewcommand{\arraystretch}{0.95}
\begin{table}[t]
\centering
\scriptsize
\caption{Circuit simulation results for completion of 1 iteration including superposition, optimization, and spin-storage update.}
\vspace{-2mm}
\label{tab:circuit}
\begin{tabular}{|cc|c|c|c|}
\hline
\multicolumn{2}{|c|}{} &
  \textbf{\begin{tabular}[c]{@{}c@{}}2 bit\end{tabular}} &
  \textbf{\begin{tabular}[c]{@{}c@{}}3 bit\end{tabular}} &
  \textbf{\begin{tabular}[c]{@{}c@{}}4 bit\end{tabular}} \\ \hline
\multicolumn{2}{|c|}{\textbf{Array Size}}                                         & $12\times36$ & $12\times48$ & $12\times60$ \\ \hline
\multicolumn{2}{|c|}{\textbf{Power {[}mW{]}}}                                     & 4.202 & 5.033 & 5.11  \\ \hline
\multicolumn{1}{|c|}{\multirow{3}{*}{\textbf{Latency {[}ns{]}}}} & \textbf{Superposition}  & 3     & 3     & 3     \\ 
\multicolumn{1}{|c|}{}                                           & \textbf{Optimization}   & 4     & 4     & 4     \\ 
\multicolumn{1}{|c|}{}                                           & \textbf{Storage Update} & 2     & 2     & 2     \\ \hline
\multicolumn{2}{|c|}{\textbf{Energy {[}pJ{]}}}                                    & 37.82 & 45.3  & 45.98 \\ \hline
\end{tabular}
\vspace{-2mm}
\end{table}

\subsection{\textbf{Architecture Simulation}} \label{sec:architecture-results}

The total latency excluding clustering with fixing simulated by PUMA simulator is presented in Fig~\ref{fig:puma_result}a. It presents the ratio of the latency with a specific maximum cluster size to that with a maximum cluster size of 12. The higher value means the slower operation. In most cases, the larger cluster size causes longer latency. 
The clustering time does not vary noticeably depending on the problem size. The latency arising from fixing the intercluster routes theoretically increases as the problem size increases, because the possible number of pairs between clusters increases quadratically while the number of clusters linearly decreases.
Based on the results presented in Fig~\ref{fig:Results}a and Fig~\ref{fig:puma_result}a, the smaller problem size is preferred if the chip can contain as many Ising macros as the number of clusters in a level. It enables mapping and solving all clusters in parallel. Otherwise, the larger size could be a better option to maximize the parallel operation. The total energy consumption excluding clustering and fixing is presented in line in Fig~\ref{fig:puma_result}a. Lower energy consumption for transferring, mapping, and processing (presented in Table~\ref{tab:circuit}) smaller $W_{D}$ makes the lower bit-precision more efficient. Table~\ref{tab:power} shows the energy consumption, excluding data transfer and mapping, to fairly compare TAXI and previous works, demonstrating that ours is the most power-efficient.

\begin{table}[!t]
%\vspace{-0.3cm}
\centering
\caption{Energy Comparison with State-of-the-Art}
\vspace{-0.2cm}
\begin{threeparttable}
\footnotesize
\resizebox{\linewidth}{!}{
\begin{tabular}{|c|c|c|c|c|}
\hline
& \cite{clustering-ising} & \cite{TSP-shimeng} & \cite{TSP-shimeng2} & \textbf{This Work} \\
\hline
\multirow{2}{*}{\textbf{Technology}} & \multirow{2}{*}{CPU} & \multirow{2}{*}{14nm FinFET} & \multirow{2}{*}{16/14nm CMOS} & 65nm CMOS +\\
                             &     &  & & SOT-MRAM \\
\hline
\textbf{Problem Size} & 101 & 1060 & 33K, 86K & 1060, 33K, 86K \\ \hline
\textbf{Energy (J)} & $1.1$& $20.08\mu$ & $\sim 20 \mu$, $\sim 45 \mu$ & $1.81\mu$, $2.67\mu$, $3.07\mu^\dagger$ \\
\hline
\end{tabular}
}
\label{tab:power}
\begin{tablenotes}
    \footnotesize
    \item $^\dagger$ Excludes mapping energy. Including mapping,  energies are $38.7\mu J$, \\ $302\mu J$, and $952\mu J$ for TSPs with sizes 1060, 33K and 86K respectively.
\end{tablenotes}
\end{threeparttable}
\vspace{-0.2cm}
\end{table}

Fig~\ref{fig:puma_result}b shows the overall latency of TAXI with a maximum cluster size of 12 along with that of Neuro-Ising~\cite{neuro-ising} and an exact solver~\cite{concorde}. We exclude CIMA ~\cite{TSP-shimeng2} from this comparison as they do not report the total latency including clustering. The bar plots display the ratio of each component contributing to the total latency. As problem size increases, the speed advantage over the exact solver gets bigger and the total latency is dominated by clustering and fixing. The biggest problem, pla85900, is solved by TAXI in 375.4s and consumes $9.51\times 10^{-4}$J, while it takes a projected time of 136 years ($1.14\times 10^{7}$ longer) and $3.82\times 10^{11}$J ($4.01\times 10^{14}$ more) for the exact solver on a single core CPU~\cite{concorde-benchmark-pla85900}.
% \vspace{-0.7mm}

\section{CONCLUSION}
We discussed existing hierarchical clustering-based Ising solvers suffer performance degradation with increasing problem size, due to search-space explosion and inherent mismatch between hierarchical execution pattern and hardware architecture data flow. To that effect, we proposed TAXI -- a hardware-algorithm co-designed TSP accelerator with Xbar-based Ising macros -- which outperforms existing Ising solvers on very large-scale TSPs. We report 8$\times$ latency improvement across 20 benchmarks over a clustering-based Ising solver\cite{neuro-ising} from TSPLib~\cite{tsplib}, up to the largest TSP with 85,900 cities. TAXI is endowed with SOT-MRAM-based in-memory Ising primitives and a hierarchical clustering algorithm. The co-design enables TAXI to solve sub-problems of large-scale TSPs on a cluster-by-cluster basis without off-macro memory access, resulting in high inter-cluster parallelism with graceful acceleration, while maintaining solution quality within an acceptable 1.2$\times$ that of exact solver~\cite{concorde} on a TSP with 85,900 cities.

\bibliographystyle{unsrt}
{\footnotesize
\bibliography{references}}%%% Uncomment this line and comment out the ``thebibliography'' section below to use the external .bib file (using bibtex).

\end{document}